# Qubit-wannabe Neural Networks

John Robert Burger[1]

Recurrent neurons, or "simulated" qubits, can store simultaneous true and false with probabilistic behaviors usually reserved for the qubits of quantum physics. Although possible to construct artificially, simulated qubits are intended to explain biological mysteries. It is shown below that they can simulate certain quantum computations and, although less potent than the qubits of quantum physics, they nevertheless are shown to significantly exceed the capabilities of classical deterministic circuits.

## Introduction

One powerful argument against the quantum mind proposition is that quantum states are based on controlled atomic particles (Quantum Mind 2011). These are thought to be unstable and thus would de-cohere before brain signaling begins. However, below it is speculated that ordinary neurons can achieve in part what qubits do. They are shown below to be more capable than ordinary bits.

It has been suggested that explicit long term memory contains associative arrays of recurrent neurons, each a ring oscillator (or multivibrator) with frequency and phase. Such elements would be microscopic, active only when called, and difficult to observe directly without destroying local activity. This led to the idea of a *simulated qubit* such that while at its highest frequency of oscillation, a deterministic true is defined; at rest a deterministic false is defined. In-between are logic levels with varying probabilities.

Envision a sphere, the top of which represents true with certainty, and the bottom of which represents false with certainty. Other points on the polar coordinate represent a continuous range of probabilities from true to false. The azimuth coordinate represents phase, which does not affect probability, but which affects certain calculations. Simulated qubits readily become controlled toggle circuits. Such controlled toggles may support the amazing feats of gifted savants (Burger 2011, 2009).

The first section below, Working with Neurons, suggests that neurons may have certain qubit-like properties even though they are based on electrical as opposed to atomic and subatomic forces. These limited qubit-like properties lead to advantages over classical neural networks. The second section Neural Networks with Quantum-like Advantages includes 1) Packing Data into a State Vector, 2) Introduction to Probability Processing. Under 2) are the topics of function classification and function satisfiability.

[1] Corresponding author: John Robert Burger, Ph D, Professor Emeritus
Address: Department of Electrical and Computer Engineering
25686 Dahlin Road
Veneta, OR 97487
e-mail: jrburger1@gmail.com

**Working with Neurons**
Neurons are nonlinear. Once triggered, they produce bursts of pulses with similar amplitudes. Each pulse tends to have a brief width but the separation can vary; the frequency of pulses within a burst may range from a few hundred hertz down to below one hertz, depending on parameters. The full purpose of these multifaceted signals has yet to be uncovered.

Fig. 1 speculates how neurons might evolve to constitute multivibrators. Frequency and therefore duty cycle may be changed by adjusting the feedback Delays (F0 and F1). Moreover, the relative phases of the output signals may be adjusted using Delays 0 and 1. All multivibrator waveforms are assumed to be harmonically related and synchronized. For example, if the low frequency is 1 Hz, then frequency can step up to perhaps 400 Hz in 1 Hz steps.

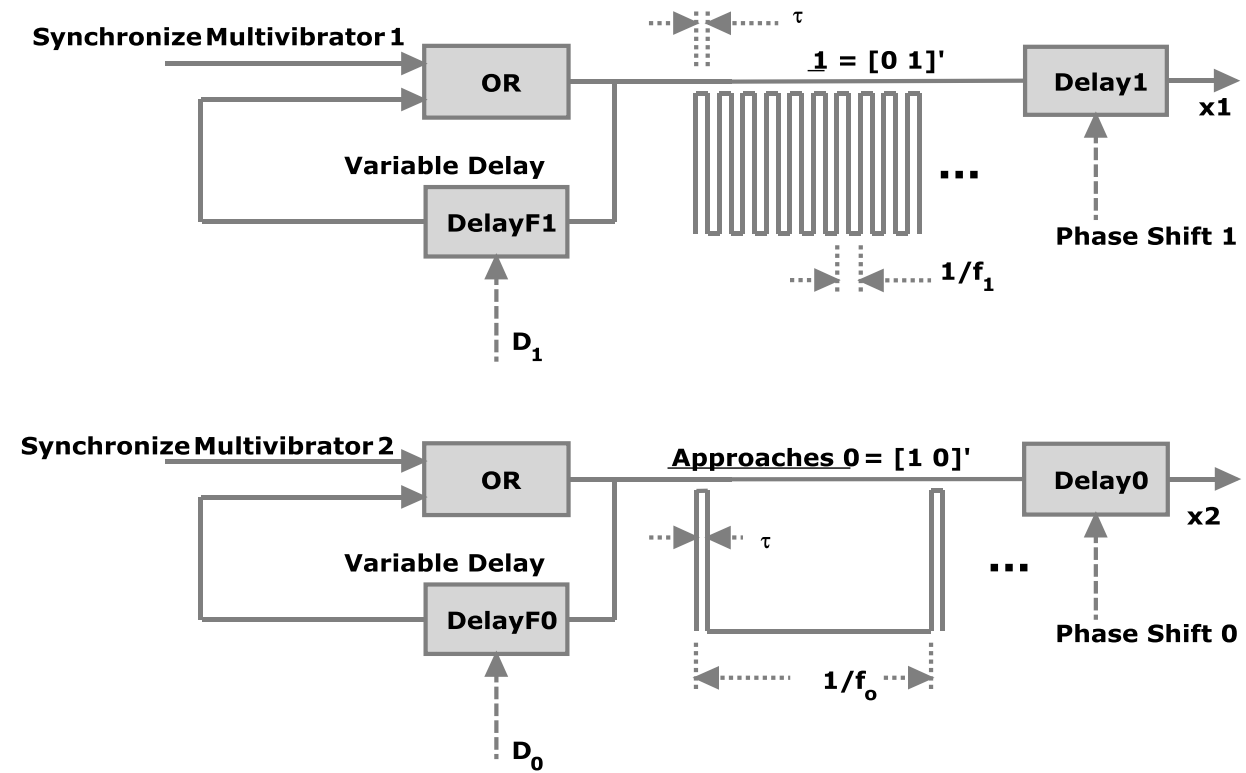

**Fig 1  Multivibrators For True (Top) and approaching False (Bottom)**

Fig. 2 illustrates a waveform that represents both true and false. A sampling window $\delta = 2\tau$ assures 100 % chance of a true at the highest frequency, assumed to be a square wave.

To observe probability of true or false a waveform of frequency $f_x$ may be sampled via a neural circuit modeled as in Fig. 3. The purpose is a properly shaped random pulse by way of a special synapse. Note that controlled toggling, which is an important application of simulated qubits, does not require a random pulse.

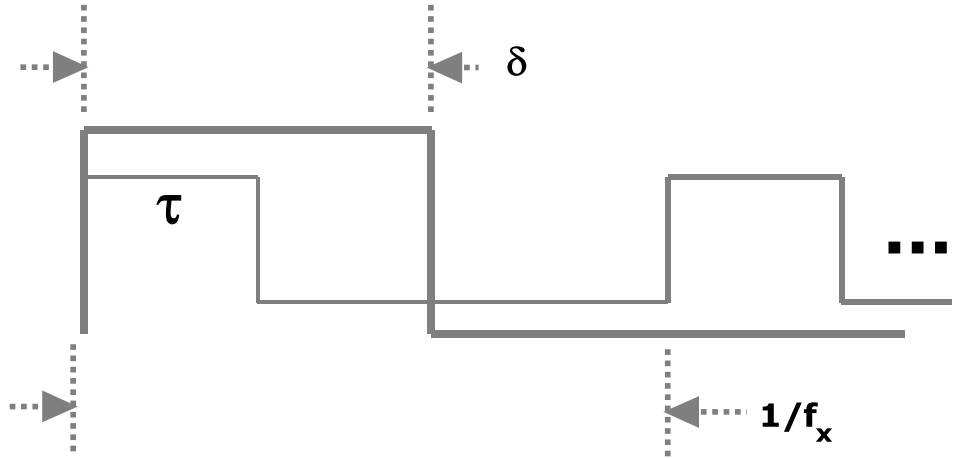

**Fig. 2 Random sampling for truth value**

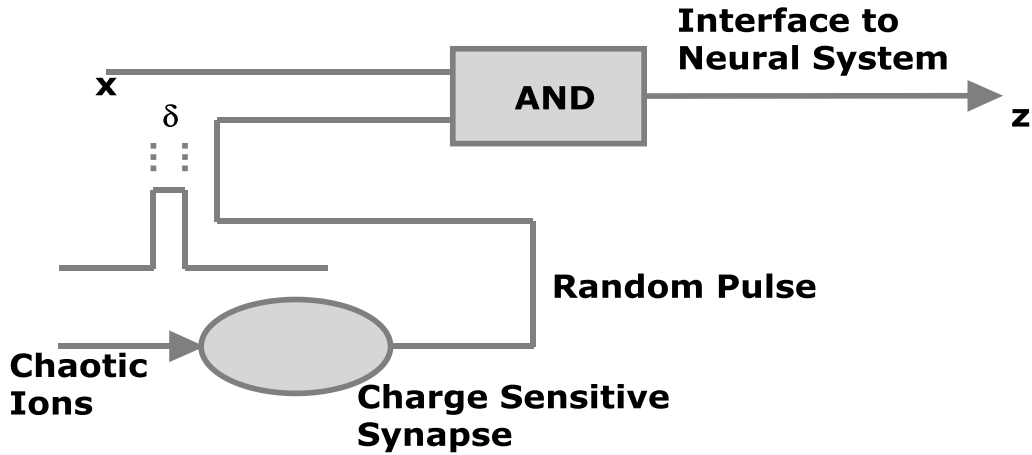

**Fig 3 Random pulse to sample combined zero and one; z is either true (an action potential) or false (at rest with no pulses)**

An intermediate frequency may be specified to be $f_x$; its period is $T_x = 1/f_x$. The width of a pulse is $\tau$. Changing frequency involves a judicious adjustment of delay for the low frequency (zero) multivibrator. The lower frequency may be specified to be $f_o \geq 0$ and the higher frequency may be specified to be $f_1$. Certainty of seeing a false occurs for $f_x = f_o = 0$. Certainty of seeing a true occurs for $f_x = f_1$ (assuming a sampling window $\delta = 2\tau$). The probability of seeing a true as an output (signal $z$) can be formulated to be:
$P_{True} \approx 2\tau/T_x = 2\tau f_x$

Therefore the probability of seeing a false is:
$P_{False} \approx 1 - 2\tau f_x$ (1)

# Delay

Biological signal delay can be regulated via the density of conductive pores (or ion channels) in an unmyelinated neural conductor and also by local ionic concentrations. Membrane capacitance is particularly significant. Special synapses are assumed to regulate delay. Fig. 4 visualizes an idealized delay segment under control of a signal labeled y.

# Analogy to Qubits

A single neural multivibrator is roughly analogous to a quantum particle or "qubit" $|\psi>$ based on two states **0**, **1**, which in vector form is **0** = (1 0)', and **1** = (0 1)' (the prime denotes a transpose and is a way to express a vertical vector on a horizontal line). A qubit is usually presented with its special bracket | > symbols, and in quantum mechanics it is generally analyzed to be a mathematically linear combination of two states:

$$|\psi> = \alpha|0> + \beta|1> \tag{2}$$

In general α and β are complex numbers. To properly conserve probability, a mathematical constraint is that:

$$|\alpha^2| + |\beta^2| = 1 \tag{3}$$

A single qubit is sometimes visualized as locating a point on a sphere as in Fig. 5 (Nielson MA and Chuang 2000, Pittenger 1999). Note that a qubit is thought of as rotating through an angle θ within the *x-z* plane. Relative phase shift involves another angle φ about the *z* axis. The result is that any qubit vector can have a positive or a negative value, or any angle in between. In particular, negative signs are permitted, for instance, either $+|\psi>$ or $-|\psi>$ is possible.

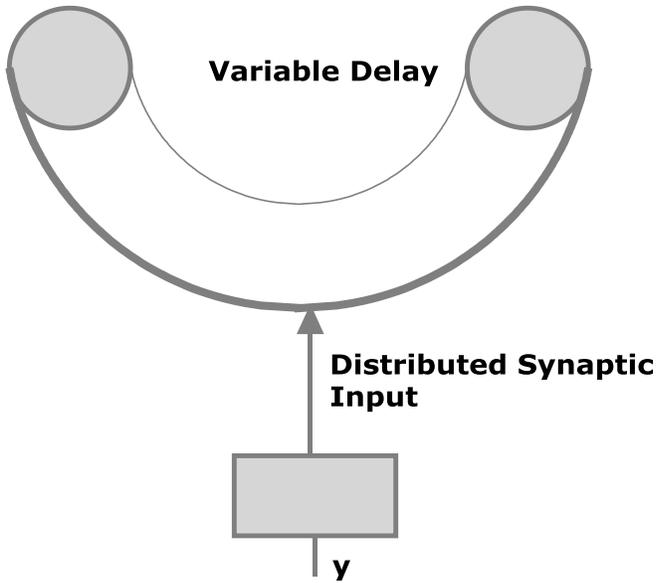

**Fig 4 Controlling Delay**

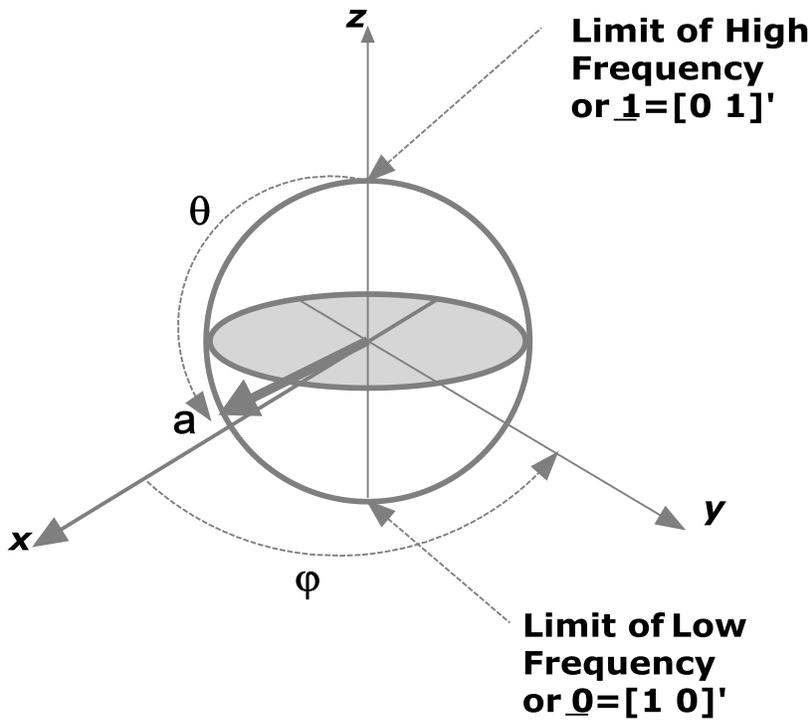

**Fig 5 Probability-phase sphere applied to neuro-multivibrators**

A sphere of probability is easy to visualize, it portrays <u>only the relative phase</u> of α relative to β, and it is good <u>for only a single qubit</u>. In a multivibrator, given an arbitrary mix of frequency and phase, an operating point can be visualized as a vector **a** = (a₁ a₂)' which could be anywhere on the sphere. Fig. 5 indicates approximately, a 50-50 mix of the two independent states, zero and one. This mix is denoted as **a** = η (1 1)', where η = $1/\sqrt{2}$ to satisfy equation (2) as developed above for a quantum qubit, that is, $(1/\sqrt{2})^2 + (1/\sqrt{2})^2 = 1$.

Suppose now that two qubits are prepared at 50 % duty cycle each. Then **a** = η (1 1)' and **b** = η (1 1)'. Upon readout, there is a 25 % chance of any given combination |00>, |01>, |10>, |11>. Thus there are four possible states. These states can be expressed using a "direct" product. That is,
**ψ** = **a b** = ($a_1$**b**  $a_2$**b**)' = ($a_1 b_1$  $a_1 b_2$  $a_2 b_1$  $a_2 b_2$)'

$$\underline{a} = \begin{bmatrix} a_1 \\ a_2 \end{bmatrix}, \quad \underline{b} = \begin{bmatrix} b_1 \\ b_2 \end{bmatrix}. \tag{4}$$

These states $a_1 b_1$, $a_1 b_2$, $a_2 b_1$, $a_2 b_2$ are variable probabilities and can be shown to be far more useful than mere binary counts, the usual result of two bits. In the special case of **a** = η (1 1)' and **b** = η (1 1)' the state vector becomes:
**ψ** = $η^2$ (1 1 1 1)'.

such that $η^2$ properly normalizes probability. For n qubits one expects $2^n$ states (Fig. 6).

Clearly there are similarities between classical multivibrators and quantum particles. Any one multivibrator can give true or false with given probabilities, as with qubits. It is easier to visualize the special case of two multivibrators with 50 % probability each. When queued with independent random sampling pulses, there is a 25 % chance of any given combination 00, 01, 10, 11; underlined symbols refers to multivibrator states, as opposed to qubit states.

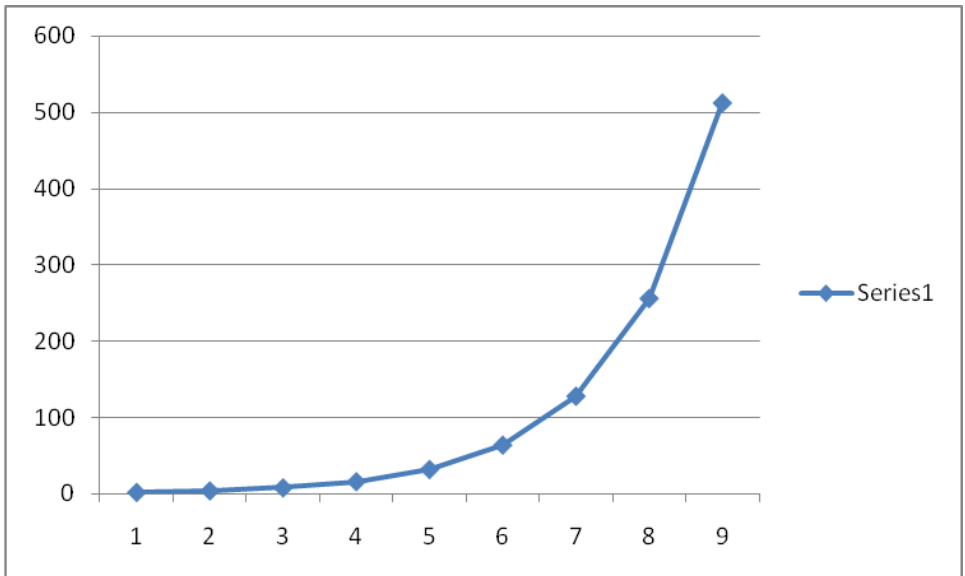

**Fig 6 Multivibrator States ($2^n$) versus n**

Burger

Unlike qubits, multivibrator probabilities take on discrete values if the (non zero) frequencies are assumed to be harmonically related. For example, if $f_o = 1$ Hz, then $f_x$ in this system is an integer; probability takes on discrete values (according to Equation 1). Note that the author places **o** as being a lower frequency and therefore, on the bottom of a sphere (usually a sphere has |**o**> at the top).

## The Phase of the One

It has not been explained yet how the phase of a one is controlled. This may be done by using an auxiliary multivibrator whose frequency is fixed at $f_1$. If the media were linear, the signals might simply add producing two tones at once, as is common mechanically. Unfortunately, neurons are far from linear. In the case of neural signals, they may combine logically, for instance, in an AND gate with a symbol as in Fig. 7, although other logical combiners are also possible.

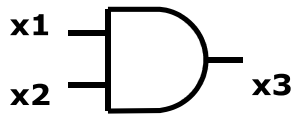

**Fig. 7 AND gate**

The AND gate is readily available in a neuron and it will preserve frequency, phase and duty cycle information. A waveform (not drawn to scale) is portrayed in Fig. 8. This waveform results by increasing the frequency of multivibrator o and keeping the frequency $f_1$ of an auxiliary multivibrator fixed.

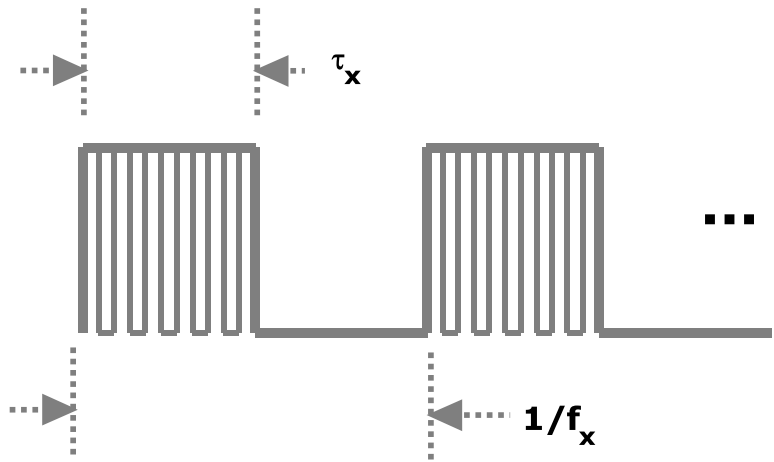

**Fig. 8 Showing No Phase Shift for a One Relative to a Zero**

## Neural Networks With Quantum-like Advantages

Neural multivibrators like those outlined above hold more information than toggle circuits or flip flops.

# 1) Packing Data Into A State Vector

Simulated qubit combinations are available for memory. For instance, binary-like information may be created with combinations of (1 0), (0 1) and η (1 1). Using (1 0), (0 1) the following probability vectors may be created: (1 0), (1 0); (1 0), (0 1); (0 1), (1 0); (0 1), (0 1), these being ordinary binary coding **0 0, 0 1, 1 0, 1 1**. In addition, using (1 0), η (1 1) the following additional codes may be created: *(1 0), η (1 1); η (1 1) (1 0)*; using *(0 1), η (1 1)* the following additional codes may be created: *(0 1), η (1 1); η (1 1), (0 1)*; and finally there is the code: *η (1 1), η (1 1)*. These codes are in addition to the 4 ordinary binary codes for *n = 2*. This implies that two simulated qubits can store *9*, which is more than *4* codes. The improvement increases exponentially with an increase in *n*

Generally the number of additional data items grows exponentially with the number n, the number of independent variables using multivibrators. By considering independent elements as being *(1 0), (0 1)* and η (1 1) there could be $2^n + n\{2^{n-1} + 2^{n-2} ... 2^1\} + 1$ codes. This calculation by the author is the binary count $2^n$ of the basic variables **0, 1**; plus the binary count with η (1 1) in place of one variable; plus the binary count with η (1 1) in place of two variables; and so on to η (1 1) in place of all variables.

The net count using multivibrators is far more than for binary coding with a mere $2^n$ codes using n bits. However, since each advanced code is probabilistic, waveforms would have to be sampled several times in order to read out faithfully the original data. This may be accomplished in this classical system by permitting several random sampling pulses and letting the data accumulate in a register for this purpose.

# 2) Introduction To Probability Processing

The availability of phase for $f_1$, indentified above as phase 1 creates interesting possibilities for multivibrators (While phase 0 is held at zero). A simulated qubit can be expressed as a combination of false and true as follows: **a** = $(a_1\ a_2)'$.

Consider multivibrator **a** that is transformed to be: **a** = $(a_1\ a_2)'$ = η (1 1)'. Mathematically this particular vector can be imagined as being analyzed as:
**a** = η (**a₁** + **a₂**)

$$\underline{a_1} = \underline{0} = \begin{bmatrix} 1 \\ 0 \end{bmatrix}, \quad \underline{a_2} = \underline{1} = \begin{bmatrix} 0 \\ 1 \end{bmatrix}.$$

(1)

In contrast, : **a** = $(a_1\ a_2)'$ = η (1 -1)' may be expressed as:

$$\underline{a_1} = \underline{0} = \begin{bmatrix} 1 \\ 0 \end{bmatrix}, \quad \underline{a_2} = -\underline{1} = \begin{bmatrix} 0 \\ -1 \end{bmatrix}.$$

(2)

Waveform for these two conditions are suggested in Fig. 9.

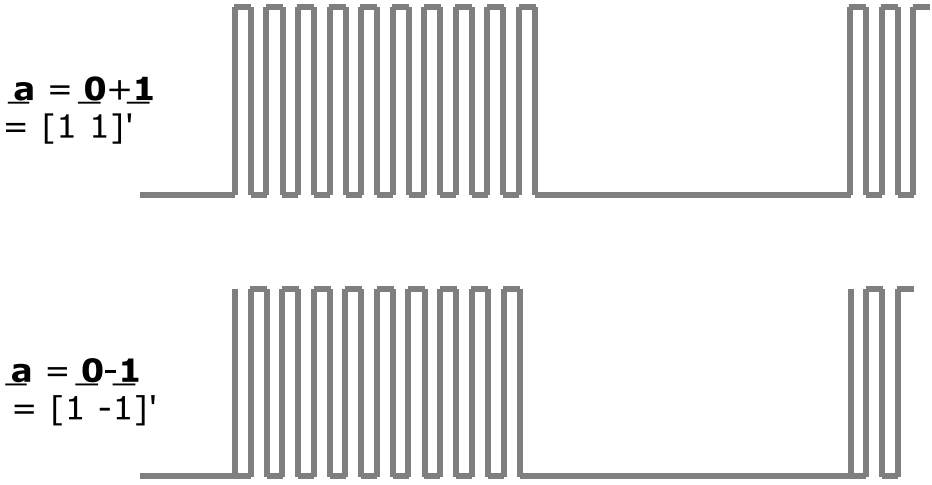

**Fig. 9 Waveform with phase shift**

# Detecting The Phase Of The 1

The problem addressed now is to distinguish a simulated qubit that is in state *η (1 1)'* from one in state *η (1 -1)'*. It is not permitted to place an oscilloscope probe directly to a multivibrator in some measuring scheme because recursive neurons are assumed too small and delicate. Rather it is desired to perform some sort of transformation on the qubit such that sampling-readout will show exactly what its phase was.

A phase shift will not affect the resultant probability after sampling, so the problem now is, how to distinguish η (1 1)' from η (1 -1)' at the output of the simulated qubit. By regulating the delays, it is possible to manipulate the frequency and the phase of a multivibrator. This in turn can be made to provide useful information after sampling, which provides a true or false. Defined is **a** = ($a_1$ $a_2$)' = η (1 1)' as a point on the sphere. Let (p q)' = (1 1)'. That is, phase shift is zero and frequency is roughly between $f_0$ and $f_1$. Next consider a transformation:

$a_1$ = *(1 - η²)* p + *η²* q = 1                                                                                        (3)

$a_2$ = *(1 - η²)* p - *η²* q = 1 − 2 *η²* = 1 − 1 = 0                               (4)

Note that *η²* = 0.5. So η (1 1)' transforms to (1 0)' which is **0** or false, determined with certainty with one observation. So it can be known if the original vector is η (1 1)'. There are given shifts in frequency and phase that do this, but identifying the shifts is beyond the scope of this analysis.

In the case of η (1 -1)', let (p q)' = (1 -1)'. Then the same transformation yields:

$a_1 = (1 - \eta^2)p + (\eta^2)(q) = 1 - 2\eta^2 = 0$         (5)

$a_2 = (1 - \eta^2)p - (\eta^2)(q) = 1$         (6)

So η (1 -1)' transforms to (0 1)' which is **1** or true, determined with certainty with one observation. So it can be known if the original vector is η (1 -1)'.

## Binary Functions Using One Simulated Qubit

A simulated qubit can be prepared to hold 0 and 1 simultaneously. Upon sampling and readout, a single truth value is presented. Assuming a binary function of a single bit, possibilities are, the function is constant: either $f(x_i) = 0$ or $f(x_i) = 1$ for $i = 1, 2$; otherwise is non-constant: either $f(x_i) = x_i$ or $f(x_i) = x_i'$, the complement of $x_i$. This information can be presented in a truth table, Table 1A:

**Table 1A The four functions of a single binary variable**

| $x_i$ | $f_0$ | $f_1$ | $f_2$ | $f_3$ |
|---|---|---|---|---|
| 0 | 0 | 0 | 1 | 1 |
| 1 | 0 | 1 | 0 | 1 |

Two functions, $f_0$ and $f_3$ are constant functions as $a_i$ varies ; the other two, $f_1$ and $f_2$ are non-constant.

A *multivibrator function* is a sequence of operations on the polar location of the probability vector and the phase (azimuth of the probability vector). The operations affect relevant variables, in this case $a_1$, $a_2$. To clarify what a multivibrator function does, refer to the inverter $f_2$ example in Table 1B.

**Table 1B Procedure to identify constant and non-constant functions**

| $x_i$ | $f_2$ | Prepared List | Tagged List | Assumed List Structure |
|---|---|---|---|---|
| 0 | 1 | 1 | -1 | $= a_1$ |
| 1 | 0 | 1 | 1 | $= a_2$ |

Imagine a qubit that is prepared to be η (1, 1); *this relates to the* prepared list in the table, in this case, $a_1 = 1$, $a_2 = 1$. After the function is applied, the *1*s in the prepared list are converted to -*1*s in those rows where the truth table for $f_2$ has a *1*, in this case the first row. Respective variables in the *assumed list structure* are then equated to the *tagged list* to create a transformed qubit that is going to be sampled and read out. In this case $a_1 = -1$, $a_2 = 1$. So the simulated qubit has changed from *η (1, 1) to η (-1, 1)*.

What has happened is that a non-constant multivibrator function, the inverter, has CHANGED the phase of the multivibrator $f_1$ such that η *(1 1)'* becomes η *(-1 1)'*. Upon transformation **-1** = (0 -1)' occurs in the multivibrator; upon sampling and readout, a true is observed, since phase does not affect the truth value. It is not difficult to show that any non-constant function will result in the observation of a true, and that any constant function will result in the observation of a false.

The following demonstration is not very impressive but it leads to interesting conclusions about simulated qubits that hold false and true simultaneously. Assume that a multivibrator function has been applied to a simulated qubit, but that the function was applied elsewhere and so the function, although implemented, is unknown. Again, it is not permitted to place an oscilloscope probe directly the multivibrator in some sort of direct observation. To help identify the function, probability processing is used as above. Assume that after applying the above probability transformation, the multivibrator is **_a_** =± *(1 0)'*. *The sample readout is false, so* it may be concluded with certainty, that the function is "constant" analogous to either $f(a_i) = 0$ or $f(a_i) = 1$ for $i = 1, 2$.

But if after transformation **_a_** =± *(0 1)'* and the sampled readout is true, then it may be concluded with certainty that the function is non-constant, and analogous to either $f(a_i) = a_i$ or $f(a_i) = a_i'$. Constancy or non-constancy is considered a global property, which can be determined by probability transformations and then observing **a** only **once** to see if it is **0** or **1**. Function classification for constancy or non-constancy usually requires evaluation at least two times, once for $x_1 = 0$ and once for $x_2 = 1$.

The above method was inspired by *Deutsch's algorithm*. For larger numbers of simulated qubits seemingly inconsequential advantages like this become significant.

## Symmetric And Antisymmetric Functions Using *n* Simulated Qubits

For a group of *n* synchronized multivibrators a function is defined to operate on each possible combination for which there is a probability after sampling and readout. For example, for *n* = 2, there is probability for each of $a_1 b_1, a_1 b_2, a_2 b_1, a_2 b_2$. What a function is defined to do is to change the phase (sign) of one or more of these terms. There are given shifts in frequency and phase that do this, but identifying the shifts is beyond the scope of this analysis.

The class of functions $f_{sa}$ (symmetric and antisymmetric functions) can be identified by their patterns in truth tables[2]. For $k > 1$ there are an even number of ones in the truth table. The limiting case of a function of one bit *(k = 1)* yields a demonstration of constancy or non-constancy as given above.

---

[2] A symmetric or antisymmetric function $f_{sa}$ of dimension k is symmetric or antisymmetric about the center of its truth table and is either symmetric or antisymmetric in each binary subdivision of $2^{k-1}$ entries of the table i=1,2...k-1 for k > 1. This class of binary functions has an even number of true entries in its truth table for k > 1.

Assume that a function in this class is made available but is unknown. A procedure for identifying an unknown $f_{sa}$ is illustrated for the case n = 2 in Table 2A. The multivibrators each have a mix of zeros and ones: $(a_1\ a_2)'$, $(b_1\ b_2)'$. The two multivibrators are prepared to have equal probability and zero phase as suggested by the *Prepared* list in the table. This list is assumed to be ordered like a binary count: $a_1\ b_1, a_1\ b_2, a_2 b_1, a_2\ b_2$. This form is inspired by the terms of a direct product $\mathbf{a} \otimes \mathbf{b}$. Probability normalization factors are omitted in the tables below in an attempt to simplify the notation.

The end result of applying the function are selected phase reversals in $a_1\ b_1, a_1\ b_2, a_2 b_1, a_2\ b_2$, as suggested by the *Tagged* list. Note that the negative signs correspond to the ones in the truth table under $f_{as}$. This is what the function accomplishes; it selectively tags the combinations $a_1\ b_1, a_1\ b_2, a_2 b_1, a_2\ b_2$ with negative signs. How a function might accomplish this is not discussed at this point.

If the combinations $a_1\ b_1, a_1\ b_2, a_2 b_1, a_2\ b_2$ are tagged with negative signs, this is equivalent to certain phase reversals for the **1**s of the multivibrators. To show this, equate the *Assumed List Structure* to the Tagged List, for example $a_1 b_1 =1$, $a_1\ b_2 = -1$, $a_2\ b_1 = -1$, $a_2\ b_2 = 1$. Begin by assuming that $a_1 = 1$. Solve the equations to obtain $a_1 = b_1 = 1$, $a_2 = b_2 = -1$. Then $(a_1\ a_2)'$, $(b_1\ b_2)' = (1\ -1)'$, $(1\ -1)'$; this in turn can be transformed to $(0\ 1)'$, $(0\ 1)'$ which upon sampling can be observed as **1 1**. There is an intimate correspondence between what is observed and the function.

**Table 2A Applying $f_{sa} = x_1 \oplus x_2$ to Multivibrator States**

| Truth Table | | | Prepared List | Tagged List | Assumed List Structure |
|---|---|---|---|---|---|
| $x_1$ | $x_2$ | $f_{as}$ | | | |
| 0 | 0 | 0 | $a_1\ b_1 = 1$ | 1 | $= a_1\ b_1$ |
| 0 | 1 | 1 | $a_1\ b_2 = 1$ | -1 | $= a_1\ b_2$ |
| 1 | 0 | 1 | $a_2\ b_1 = 1$ | -1 | $= a_2\ b_1$ |
| 1 | 1 | 0 | $a_2\ b_2 = 1$ | 1 | $= a_2\ b_2$ |

To recap, a function of two binary variables has the truth table "0 1 1 0" which is symmetric (about its center). Prepared multivibrators $(a_1\ a_2)'$, $(b_1\ b_2)' = (1\ 1)'$, $(1\ 1)'$ are tagged as in the above table and transformed to become $(1\ -1)'$, $(1\ -1)'$; This, in turn, transforms to $(0\ 1)'$, $(0\ 1)'$ which can be observed deterministically to be **1 1**.

If a negative sign were associated with the **1 1** it would not be seen, because sampling does not respond to phase. For example, a function with the truth table "1 0 0 1" is symmetric (about its center). Prepared multivibrators can be transformed as in the above table to become $(-1\ 1)'$, $(1\ -1)'$; This, in turn, transforms to $-(0\ 1)'$, $(0\ 1)'$ but this is still observed to be **1 1**.

Functions such as 0 1 1 0 and 1 0 0 1 are termed *complementary* functions. Complementary functions give the same sampled output. Thus a result of **1 1** identifies a given function "0 1 1 0" or its complement "1 0 0 1" within the class of symmetric and antisymmetric functions.

As another example of this, a function with a truth table 0 0 1 1 is antisymmetric; it can be made to convert (1 1)', (1 1)' to (1 -1)', (1 1)' in Table 2B. This transforms to **1 0** which identifies the function as being either 0 0 1 1 or 1 1 0 0.

**Table 2B Applying $f_{sa} = a_2$ to Multivibrator States**

| Truth Table $x_1$ $x_2$ $f_{as}$ | Prepared List | Tagged List | Assumed List Structure |
|---|---|---|---|
| 0  0   0 | 1 | 1 | $= a_1 b_1$ |
| 0  1   0 | 1 | 1 | $= a_1 b_2$ |
| 1  0   1 | 1 | -1 | $= a_2 b_1$ |
| 1  1   1 | 1 | -1 | $= a_2 b_2$ |

It can be concluded that by observing the values and the phases for $n$ multivibrators, one of $2^n$ symmetric and antisymmetric functions can be identified to within a complement with certainty, using only one observation. Normally a function of n binary variables would require up to $2^n$ calls to the function to fill in its truth table. So if an unknown function within the class $f_{sa}$ is applied to multivibrators, it can be identified to within a complement using procedures as above.

## 3) The Satisfiability Problem

An interesting class of binary functions $f_d$ may be defined on integers $k$, $0 \leq k \leq 2^n-1$, so that $f_d(k) = 0$ for all $k$ except for $k = k_o$, and $f_d(k_o) = 1$. These are known as *decoding* functions. Assume for the moment that $f_d$ is given but that what satisfies it, $k_o$ is unknown. If one searches classically, evaluating $f_d(k)$ until one finds $k_o$ one might have to make as many as $2^n$ evaluations of $f_d(k)$.

When a binary function is known, it is not always easy to discover what satisfies it, especially for complicated or strangely coded functions. Finding what satisfies a given binary function such as $f_d(k)$ is known as a *satisfiability problem*.

Here is a simple example in which the normalization factors have been omitted in order to focus on the vectors. Consider a neural logic system that provides the binary function in Table 3A. Note that the truth table is all zeros except for a single one, so it is a decoding function; it satisfies the condition that only one code satisfies the function.

**Table 3A Decoding Function $f_d = x_1' \, x_2$**

| Truth Table | | | Prepared List | Tagged List | Assumed List Structure |
|---|---|---|---|---|---|
| $x_1$ | $x_2$ | $f_{as}$ | | | |
| 0 | 0 | 0 | 1 | 1 | $a_1 \, b_1$ |
| 0 | 1 | 1 | 1 | -1 | $a_1 \, b_2$ |
| 1 | 0 | 0 | 1 | 1 | $a_2 b_1$ |
| 1 | 1 | 0 | 1 | 1 | $a_2 \, b_2$ |

Simulated qubits, once sampled, can provide arbitrary combinations of true and false. Although a mystery, the possible combinations are imagined to exist prior to sampling. A multivibrator function is assumed to tag items with minus signs where the function is satisfied, as suggested in the table. How a multivibrator function might accomplish the tagging is not speculated upon at this point.

In an attempt to discover which $a_i \, b_j$ term is the -1, a transformation is going to be applied to the multivibrators involved, prior to sampling and readout. The list is permutated using an **H** matrix of **H** matrices, or $\mathbf{H}^{\otimes 2}$ giving the result *Mod1* shown in table 3B. Note that:

$$\mathbf{H} = \frac{1}{\sqrt{2}} \begin{bmatrix} 1 & 1 \\ 1 & -1 \end{bmatrix}$$

This list is modified again with a transformation that reverses the phase of the first entry $a_1 \, b_1$ as shown in the right of the table under Mod2.

**Table 3B Decoding Function $f_d = x_1' \, x_2$**

| | | Mod1 | Mod2 |
|---|---|---|---|
| 1 | $a_1 \, b_1$ | 1 | -1 |
| -1 | $a_1 \, b_2$ | 1 | 1 |
| 1 | $a_2 b_1$ | -1 | -1 |
| 1 | $a_2 \, b_2$ | 1 | 1 |

The sign of the $a_1 \, b_1$ term is changed so that there are an even number of negative signs in the list under Mod2 in the table. At this point a result can be identified.

A transformation is defined by solving the four equations in the table, givng **a** = (1  1)' and **b** = (-1  1)'. The probabilities can be transformed as above to show **a b** = -**0 1**. The minus sign will not be observed after sampling, giving an output of **0 1** or position two from the top. This is where the original *1* was in the original Table 3A for this function.

This method of satisfiability analysis was inspired by *Grover's algorithm*. It can work for more than two simulated qubits, but an iteration is necessary, about $2^{n/2}$ times. That is, to summarize, 1) use the function to tag the list, 2) apply $\mathbf{H}^{\otimes n}$ to the tagged list, 3) reverse the phase of the first term in the list, and 4) solve for an updated result **a b** filled with fractions. This last step is equivalent to applying $\mathbf{H}^{\otimes n}$ again. When the iterations are completed, sampling may take place to provide with good probability a solution to the problem of what satisfies the given function. The solution must be checked by substitution, and if it fails, the method is repeated.

This method is a vast improvement over having to run a function for up to $2^n$ times (each possible input combination) trying to discover what satisfies it.

## Conclusions

This chapter explores possibilities for transforming the states of simulated qubits. These transformations may apply to neurons, although exactly how is a frontier. No one has proved that simulated qubits and their multivibrator functions are impossible, yet no one can say how such functions might be implemented or used to advantage. Clearly there are potential advantages.

Multivibrator functions modify the position of the state vector on the sphere. Fundamental to multivibrator functions are transforms, introduced above, that permit distinguishing state *η (1 1)'* from state *η (1 -1)'* upon sampling and readout. Using such transforms, it was shown that if a multivibrator function is applied but is somehow unknown, transforms permits function classification. This was shown for symmetric and antisymmetric functions involving *n* simulated qubits, for which only a single readout classified an unknown function to within a complement. (Normally up to $2^n$ evaluations would be necessary). It is possible that such a function might be applied to simulated qubits within a brain, and that identification of the function upon readout represents an important mental realization.

An interesting and mind-boggling concept is that there is a transformation that will identify, upon sampling and readout, the binary code that satisfies a given function. If the function is complex, it could be far from obvious what satisfies it. For example, there are Boolean operations that serve to define the prime factors a large number. Using brute force, it would take many trials to find the prime factors, since a great many prime number would have to be tested. But if the function can be implemented on simulated qubits, there is a way to find what satisfies it using only one, or at least only a few sample-readout operations.

In the case of a decoding function, for which there is only one binary code that satisfies it, the function needs to be evaluated only about $2^{n/2}$ times. Ordinarily it would require up to $2^n$ evaluations. So there may be significant advantages for larger *n*. Fanciful but possible is that neurons have a way to determine what satisfies important functions of biological interest.

This chapter brings up the possibility that simulated qubits can store a great deal more coded information that ordinary long term memory circuits. The increase occurs because simulated qubits can store true and false simultaneously, while having phase available. For example, the brain might represent an attribute of a mental image and also encode its strength (with frequency) and its color (with phase). To give a computer-like example using (1 0)', (0 1)' and η (1 1)', the number of codes that can be stored increases exponentially with *n* and becomes far more than a mere $2^n$, which is all that *n* bits provide.

Probability upon readout is no problem in practice, at least not classically, because it is easy to read a simulated qubit more than once, and accumulate a result. Memory is important, so options for increasing memory density, or the information per memory circuit, cannot be dismissed without due consideration.